# ‘Just another field theory’ approach to $\mathcal{N} = 1$ Super Yang-Mills and the origin of intrinsic SuperGeometry

Konstantinos Koutrolikos[1]

*Department of Physics, University of Maryland,*
*College Park, MD 20742-4111, USA*

ABSTRACT

We present a non-geometric derivation of $\mathcal{N}$=1 Super Yang-Mills by focusing on the consistency of interactions that extend the free vector supermultiplet rather than assuming gauge invariance under extended symmetries. By utilizing a superspace first-order description, the theory is given in closed form as a third-order polynomial which includes a single cubic interaction term instead of an infinite series, thus eliminating the need for a special gauge. The geometrical interpretation of the theory emerges, as opposed to being presupposed.

[1] koutrol@umd.edu

## 1 Introduction

The relation between geometry and theoretical physics has been extremely successful and fruitful across a wide range of applications, including string theory, non-linear $\sigma$-models, topological field theories, Yang-Mills theory, and of course gravity. As a result, understanding the geometric meaning of the dynamics and interactions of a physical system has become a preferred method of analysis. The prototypical example is gravity where the geometrical description of spacetime in terms of differentiable manifolds determines (*self*)interactions and generates the non-linearity of the theory to all orders. In classical gauge theories, the gauge fields are connections on a principal G-bundle and the physically observable field strengths correspond to the curvatures associated with these connections, while matter corresponds to associated vector bundles. Quantization of these theories via path integrals amounts to integrating over the space of connections on principal fiber bundles.

Nevertheless, despite the significant advances made by the geometric approach, it is still necessary to develop additional methods to fully grasp the underlying structure and behavior of these theories. For example, one may argue that while the geometrical interpretation of gauge theories, like QCD, has led to many useful insights and topological aspects of them, it has not helped us completely solve these theories. Moreover, in the case of gravity, the conventional geometrical approach may obscure the path towards the quantization of the theory as one has to define up front the physical meaning of quantum geometry, fluctuating topology, etc. Another compelling example that illustrates the usefulness of a non-geometric approach to constructing interacting gauge theories can be found in higher spin gauge theories, where the absence of an underlying geometry[2] leaves us without a deep understanding of their interactions.

The existence of a non-geometric methodology was demonstrated and pioneered [3–8] in the derivation of gravity as the interacting theory of massless spin-2[3] field in ordinary Minkowski spacetime. This is a field theoretic approach which treats gauge theories "*just as any other field theory*" and it focuses on examining the consistency of modifications to the linearized, free theory. In particular, for gravity one can generate the infinite series of non-linear, self-interaction terms by following this well-known argument: The linearized theory generates a quadratic energy-momentum tensor that can play the role of source for interactions. This induces a cubic interaction term in the Lagrangian which will contribute a new cubic contribution to the energy-momentum tensor and therefore a new source. Consequently, a quartic interaction term has to be introduced in the Lagrangian and so on. Once initiated, this iterative process must continue to all orders[4]. Therefore, the theory is either left free or it acquires an infinite series of self-interactions. Amazingly, all these infinite terms can be resumed, if derived properly [14], to yield the full nonlinear Einstein equations.

[2]A detailed introduction to the algebra and geometry of differential operators and jet bundles can be found in [1, 2].

[3]For similar considerations of interacting fields with definite spin values see [9–13].

[4]The source on the right hand side of the free, massless, spin-2 equation of motion must be conserved as the left hand side of the equation is divergentless due to linearized gauge redundancy. This conservation only holds for the full infinite series.

Moreover, consistency of interactions with matter leads to universal coupling and therefore to the equivalence principle which is the origin of the geometrical interpretation of the theory. All matter moves in an effective Riemannian manifold while the initial Minkowski background disappears. The existence of this geometrical interpretation of gravity (and similarly for Yang-Mills) is undeniably remarkable and profoundly deep. However, it requires careful consideration as to whether it is merely a low spin accident —a coincidence that reflects the properties and gauge redundancy intrinsic to free spin-2 (or spin-1) fields— or a feature that must be expected from higher spin gauge theories.

For manifestly supersymmetric theories, the development of a similarly inspired non-geometrical approach to constructing interacting gauge theories is even more pivotal. This is primarily due to the fact that the conventional geometrical construction of Super Yang-Mills [15] and Supergravity [16, 17] relies on introducing superconnection superfields, which encompass numerous component fields that need to be eliminated. This inevitably entails the laborious task of identifying a consistent set of constraints, solving these constraints and the associated Bianchi identities and then expressing the potential superfields in terms of unconstrained prepotentials. On the contrary, the non-geometric approach to deriving these theories is to consider the consistent interactions of superspin $\mathsf{Y} = \frac{1}{2}$ (*vector*) and $\mathsf{Y} = \frac{3}{2}$ (*supergravity*) supermultiplets respectively. The advantage lies in the fact that the unconstrained prepotential superfields are inherently incorporated into the theory. They correspond to the unconstrained superfields participating in the description of the linear theories. Hence, there is no need to impose[5] and solve complex constraints. Moreover, the appropriate geometrical interpretation (if any) of the prepotentials[6] is expected to emerge naturally, rather than being assumed from the outset.

The scope of this paper is to demonstrate the existence of a non-geometric approach, as described above, in deriving interacting gauge theories within Superspace. In particular, we showcase the derivation of $\mathcal{N} = 1$ Super Yang-Mills theory as the consistent self-interacting system extending from the initial free, massless, vector supermultiplet. The theory is presented in a simple, closed form with a single cubic interaction rather than an infinite series of self-interactions and without the need to introduce a special gauge. This is made possible due to the first order formalism developed in [26, 27]. Although this methodology stresses consistency of interactions rather than gauge invariance, the complete theory is invariant under an extended group of symmetries despite the fact that this was not initially required. This is the big difference between the derivation presented here and those approaches that rely on the requirement of invariance.

[5]For completeness, we mention the "*coupling to matter*" methodology of constructing SYM [18] and SUGRA [19, 20]. In this approach one assumes the anticipated geometrical structures and after postulating how matter supermultiplets (such as chiral ones) transform in the presence of geometry, the existence of unconstrained gauge superfields —acting as *bridges*— and their redundancies are inferred. These can be used to construct additional geometrical objects such as covariant derivatives and curvatures as well as invariant actions. This approach has been used [21–23] successfully to generalize the notion of linearized superdiffeomorphism to higher spin transformations of matter multiplets and construct non-trivial interactions between matter and arbitrary high superspin multiplets.

[6]The prepotentials of $\mathcal{N} = 1$ SYM [24, 25] take values in the coset $G_C/G$, where $G_C$ is the complexified group G.

Specifically, the use of (*i*) first-order description [26] of the vector supermultiplet featuring spinorial connection-like superfields which are à priori independent from the gauge superfield and (*ii*) suitable superfield variables, shortcuts the otherwise infinite iterative process of constructing higher-order interaction terms into a single step yielding just one cubic term. The '*resummation*' process becomes trivial and the action acquires a simple polynomial form.

In this form of the action, the initial vector supermultiplets gauge superfields become imperceptible and they are replaced instead by more natural (and fundamental) gauge superfields. This serves as the origin of the theory's geometric interpretation, as it generates all the familiar geometrical objects such as, superfield strength, superspace covariant derivatives and Bianchi identities. It is worth emphasizing that throughout this process there is no need to consider any set of (*conventional*) constraints or to go to a special gauge.

## 2 SYM as interacting vector supermultiplets

In [26, 27] a superspace *first-order* formulation of supersymmetric gauge theories was presented. Following the usual first-order philosophy, this method introduces additional and independent auxiliary superfields that play the role of connections for new (often algebraic) local symmetries. These new symmetries allow the relaxation of linear gauge redundancies and lead to the construction of simpler gauge invariant actions. The dynamics of these connection-like superfields are governed by algebraic equations of motion, thus one can integrate them out reproducing the known "*second order*" superspace actions. The benefit of this formulation is the simplification of constructing consistent interactions which now take a polynomial form. This is a desired property for understanding non-perturbative effects.

The first order description of the vector (*Maxwell*) supermultiplet, with superspin $\mathsf{Y} = \frac{1}{2}$ requires a pair of auxiliary, spinorial, superfields $\Omega_\alpha$ and $\mathcal{W}_\alpha$. The superspace action takes the simple form[7]:

$$S_o = \int d^8z \left\{ \mathcal{W}^\alpha \Omega_\alpha + \mathcal{W}^\alpha \mathrm{D}_\alpha V - 2c\, \Omega^\alpha \bar{\mathrm{D}}^2 \Omega_\alpha \right\} + c.c. \tag{1}$$

and it is invariant under the following linear gauge transformations:

$$\delta \mathcal{W}_\alpha = 0\ ,\ \delta \Omega_\alpha = -\mathrm{D}_\alpha \bar{\mathrm{D}}^2 L\ ,\ \delta V = \bar{\mathrm{D}}^2 L + \mathrm{D}^2 \bar{L}\ . \tag{2}$$

Deriving the equations of motion and integrating out superfields $\mathcal{W}_\alpha$, $\Omega_\alpha$

$$\mathcal{E}^{(\mathcal{W})}_\alpha := \Omega_\alpha + \mathrm{D}_\alpha V = 0\ ,\ \mathcal{E}^{(\Omega)}_\alpha := \mathcal{W}_\alpha - 4c\, \bar{\mathrm{D}}^2 \Omega_\alpha = 0\ ,\ \mathcal{E}^{(V)} := -\mathrm{D}^\alpha \mathcal{W}_\alpha - \bar{\mathrm{D}}^{\dot\alpha} \bar{\mathcal{W}}_{\dot\alpha} = 0 \tag{3}$$

[7]This is exactly the same action as in [26] but written in different variables: $V = \mathcal{V} + \bar{\mathcal{V}},\ \mathcal{W}_\alpha = \mathtt{W}_\alpha + 2c\, \bar{\mathrm{D}}^2 \Omega_\alpha$.

we find the familiar vector supermultiplet action:

$$S_o = 4c \int d^8z \left\{ V \mathrm{D}^\alpha \bar{\mathrm{D}}^2 \mathrm{D}_\alpha V \right\} \ . \tag{4}$$

Notice that superfield $\mathcal{W}_\alpha$ on-shell produces the linearized superfield strength $\mathrm{W}_\alpha$, while $\Omega_\alpha$ relates to the linearized superconnection [28] $\Gamma_\alpha$

$$\mathcal{W}_\alpha\Big|_{\delta S_o=0} \propto \mathrm{W}_\alpha = \bar{\mathrm{D}}^2\mathrm{D}_\alpha V \ , \ \ \Omega_\alpha\Big|_{\delta S_o=0} \propto \Gamma_\alpha = \mathrm{D}_\alpha V \, . \tag{5}$$

Substitution into $\mathcal{E}^{(V)} = 0$ and using the identity $\mathrm{D}^\alpha \mathcal{W}_\alpha = \bar{\mathrm{D}}^{\dot{\alpha}} \bar{\mathcal{W}}_{\dot{\alpha}}$, yields the linear equation

$$\mathcal{E} \equiv \mathrm{D}^\alpha \mathcal{W}_\alpha = \mathrm{D}^\gamma \bar{\mathrm{D}}^2 \mathrm{D}_\gamma V = 0 \, . \tag{6}$$

We now demand that the right-hand side of this equation be augmented to include a source $\mathrm{J}$, which accounts for interactions with matter or other gauge supermultiplets. Ensuring the consistency of such a modification, if it exists, necessitates that $\mathrm{J}$ satisfies on-shell the conservation equations $\mathrm{D}^2\mathrm{J} = 0 = \bar{\mathrm{D}}^2\mathrm{J}$, reflecting the presence of linear gauge redundancy. Following the standard field-theoretic approach of investigating such interactions, one would search perturbatively for non-trivial solutions to these consistency conditions. This procedure, as described in the introduction, will lead to an infinite series of terms that (a.) need to be resumed and (b.) produce the non-polynomial self interactions of SYM and its coupling to matter. Of course, such infinite summations have never been performed. Instead they are replaced by arguments based on the expected group G-covariance of the final result and other related geometric statements that circumvent the entire process. As we will demonstrate, by adopting the first-order description (1), the infinite series collapses to a single cube interaction term and the usual geometric notions of G-covariance emerge without being postulated.

A particularly straightforward method of generating appropriately conserved quantities[8] is by taking the variational derivative of (1) with respect to a fictitious supersymmetric connection $\Psi_\alpha$ which covariantizes the superspace action by replacing supersymmetric derivatives $\mathrm{D}_\alpha$ with covariant supersymmetric derivatives $\mathcal{D}_\alpha = \mathrm{D}_\alpha + \Psi_\alpha$. This is just a mathematical recipe which can be used to quickly generate relevant terms and has no deeper (geometric) interpretation. Such terms can be easily constructed in different ways. With that in mind, we examine the following cubic interaction

$$S_1 = \mathrm{g} \int d^8z \left\{ \mathcal{W}^\alpha V \Omega_\alpha \right\} + c.c. \tag{7}$$

which is a member of the $\mathcal{J}^\alpha \Omega_\alpha$ family of terms and is generated by a $\mathcal{J}_\alpha$ defined as follows

$$\mathcal{J}_\alpha := \lim_{\Psi^\alpha \to 0} \frac{\delta S_o(\mathrm{D} \to \mathcal{D})}{\delta \Psi^\alpha} = \mathcal{W}_\alpha V \tag{8}$$

[8]One can calculate the energy momentum tensor of a theory by taking the variatonal derivative of the action with respect to a fictitious metric that covariantizes the action and at the end take the limit to its background value.

Given this interaction, we must check that it satisfies two necessary consistency requirements: (1.) it is (on-shell) gauge invariant and (2.) it is non-trivial meaning that it can not be absorbed by field redefinitions. It is straightforward to verify that on-shell, $S_1$ is invariant under gauge transformations (2). However, it fails the non-triviality test as the interaction term itself vanishes on-shell. This means that $S_1$ can be written in a form proportional to equations of motion (3), hence it is not a proper, non-trivial interaction and it can be absorbed by some redefinition of the superfields:

$$S_1 = g\int d^8z \left\{ \mathcal{W}^\alpha V \Omega_\alpha \ + c.c. \right\} = g\int d^8z \left\{ \mathcal{W}^\alpha V \big(\mathcal{E}^{(\Omega)}_\alpha - \mathrm{D}_\alpha V\big) + c.c. \right\}$$
$$= g\int d^8z \left\{ \mathcal{W}^\alpha V \mathcal{E}^{(\Omega)}_\alpha + \bar{\mathcal{W}}^{\dot\alpha} V \bar{\mathcal{E}}^{(\Omega)}_{\dot\alpha} + \frac{1}{2}\mathcal{E}^{(V)} V^2 \right\} \tag{9}$$

This failure stems from the property of $V$ to commute with its derivative $\mathrm{D}_\alpha V$ which enables the expression $V(\mathrm{D}_\alpha V)$ to be written as a total derivative $\frac{1}{2}\mathrm{D}_\alpha(V^2)$. This simple fact is at the core of *super*Maxwell theory's linearity and the absence of self-interactions for the photon. Understanding this allows for a straightforward resolution by considering many copies of the vector multiplet $V^I$, $I = 1, 2, ...n$ or equivalently promoting superfield $V$ to a matrix superfield $\mathbf{V}$ such that $[\mathbf{V}\,,\, \mathrm{D}_\alpha \mathbf{V}] \neq 0$. Therefore, we consider the following interacting theory

$$S = \mathrm{Tr}\left[ \int d^8z \left\{ \boldsymbol{\mathcal{W}}^\alpha \boldsymbol{\Omega}_\alpha + \boldsymbol{\mathcal{W}}^\alpha\, \mathrm{D}_\alpha \mathbf{V} - 2c\, \boldsymbol{\Omega}^\alpha\, \bar{\mathrm{D}}^2 \boldsymbol{\Omega}_\alpha + g\, \boldsymbol{\mathcal{W}}^\alpha\, \mathbf{V}\, \boldsymbol{\Omega}_\alpha \right\}\right] + c.c. \tag{10}$$

We will show, that (10) constitutes the full $\mathcal{N} = 1$ Super Yang-Mills theory. Observe that this action contains a single cubic interaction and no higher-order terms are necessary. The infinite iterations end at this cubic term since it lacks any derivatives, thus it remains independent of the fictitious connection $\Psi_\alpha$ and as a result $\mathcal{J}_\alpha$ (8) acquires no higher order corrections. Moreover, action (10) provides a first-order description of the superspace SYM action by identifying the natural variable $\mathbf{U} = \mathbb{1} + g\,\mathbf{V}$. By collecting terms, equation (10) can be expressed as follows

$$S = \mathrm{Tr}\left[ \int d^8z \left\{ \boldsymbol{\mathcal{W}}^\alpha\, (\mathbb{1} + g\,\mathbf{V})\, \boldsymbol{\Omega}_\alpha + \frac{1}{g}\, \boldsymbol{\mathcal{W}}^\alpha\, \mathrm{D}_\alpha(\mathbb{1} + g\,\mathbf{V}) - 2c\, \boldsymbol{\Omega}^\alpha\, \bar{\mathrm{D}}^2 \boldsymbol{\Omega}_\alpha \right\}\right] + c.c. \tag{11}$$

suggesting the introduction of a more appropriate set of variables

$$\mathbf{U} = \mathbb{1} + g\,\mathbf{V}\,,\ \ \mathscr{W}_\alpha = \frac{1}{g}\, \boldsymbol{\mathcal{W}}_\alpha\,,\ \ \omega_\alpha = g\, \boldsymbol{\Omega}_\alpha\,. \tag{12}$$

In these variables, the action takes the form:

$$S = \mathrm{Tr}\left[ \int d^8z \left\{ \mathscr{W}^\alpha\, \mathbf{U}\, \omega_\alpha + \mathscr{W}^\alpha\, \mathrm{D}_\alpha \mathbf{U} - \frac{2c}{g^2}\, \omega^\alpha\, \bar{\mathrm{D}}^2 \omega_\alpha \right\}\right] + c.c. \tag{13}$$

Another important observation is that within this set of variables, superfield $\mathbf{V}$ does not mark the start of an expansion series, rather it represents the complete deviation of superfield $\mathbf{U}$ from the *background* value $\mathbb{1}$. This is in contrast with the conventional choice of gauge superfield $\boldsymbol{\mathsf{V}}$: $\mathbf{U} = e^{(g\boldsymbol{\mathsf{V}})} = \mathbb{1} + g\,\boldsymbol{\mathsf{V}} + \frac{g^2}{2!}\boldsymbol{\mathsf{V}}^2 + \ldots$ . The two are related by

$$\mathbf{V} = \frac{e^{(g\boldsymbol{\mathsf{V}})} - \mathbb{1}}{g} \tag{14}$$

and coincide at the linearized limit $g \to 0$. This change of variables is related to the *minimal homotopy* class of maps introduced in [29–31]. Additionally, it is used in [32, 33] as a convenient calculation tool for gauge superfields in *Projective* Superspace. The fact that the theory can be written purely in terms of $\mathbf{U}$, without any background dependence, suggests that the more fundamental object may not be $\mathbf{V}$ itself (or any other associated variable such as $\boldsymbol{\mathsf{V}}$), but rather $\mathbf{U}$.

We now check that (13) is indeed the $\mathcal{N} = 1$ super Yang-Mills action and produces the correct dynamics. The equations of motion generated by (13) are:

$$\mathscr{E}^{(\mathscr{W})}_\alpha := \mathbf{U}\,\omega_\alpha + \mathrm{D}_\alpha\mathbf{U} = 0\,, \tag{15a}$$

$$\mathscr{E}^{(\omega)}_\alpha := \mathscr{W}_\alpha\,\mathbf{U} - \frac{4c}{g^2}\,\bar{\mathrm{D}}^2\omega_\alpha = 0\,, \tag{15b}$$

$$\mathscr{E}^{(\mathbf{U})} := \omega^\alpha\,\mathscr{W}_\alpha - \mathrm{D}^\alpha\mathscr{W}_\alpha + h.c. = 0\,. \tag{15c}$$

Equations (15a) and (15b) are algebraic and can be solved for $\omega_\alpha$ and $\mathscr{W}_\alpha$ respectively:

$$\omega_\alpha = -\,\mathbf{U}^{-1}\,(\mathrm{D}_\alpha\mathbf{U})\,, \tag{16a}$$

$$\mathrm{W}_\alpha := \mathscr{W}_\alpha\,\mathbf{U} = \frac{4c}{g^2}\,\bar{\mathrm{D}}^2\omega_\alpha = \frac{4c}{g^2}\,\bar{\mathrm{D}}^2\Big[\mathbf{U}^{-1}\,(\mathrm{D}_\alpha\mathbf{U})\Big] \tag{16b}$$

The solution for $\omega_\alpha$ is the supersymmetric Maurer-Cartan term (16a) and as we will see it will play the role of the connection for the superspace covariant derivative that will emerge. The solution for $\mathscr{W}_\alpha$ introduces $\mathrm{W}_\alpha$ (16b), the chiral SYM superfield strength. Within these solutions, the familiar infinite non-linearities of SYM appear due to the inverse matrix $\mathbf{U}^{-1} = (\mathbb{1} + g\,\mathbf{V})^{-1}$. The presence of the background $\mathbb{1}$ ensures the existence of $\mathbf{U}^{-1}$, at least at infinity where $\mathbf{V}$ vanishes. Using these solutions, equation (15c) takes the form:

$$\begin{aligned}
0 = \mathscr{E}^{(\mathbf{U})} &= -\Big(\mathrm{D}^\alpha + \mathbf{U}^{-1}\,(\mathrm{D}^\alpha\mathbf{U})\Big)\Big[\mathrm{W}_\alpha\,\mathbf{U}^{-1}\Big] + h.c. \\
&= -\Big(\mathrm{D}^\alpha\,\mathrm{W}_\alpha\Big)\,\mathbf{U}^{-1} - \mathrm{W}^\alpha\Big(\mathrm{D}_\alpha\mathbf{U}^{-1}\Big) - \mathbf{U}^{-1}\,(\mathrm{D}^\alpha\mathbf{U})\,\mathrm{W}_\alpha\,\mathbf{U}^{-1} + h.c. \\
&= -\Big(\mathrm{D}^\alpha\,\mathrm{W}_\alpha + \mathbf{U}^{-1}\,(\mathrm{D}^\alpha\mathbf{U})\,\mathrm{W}_\alpha\Big)\,\mathbf{U}^{-1} - \mathrm{W}^\alpha\Big(\mathrm{D}_\alpha\mathbf{U}^{-1}\Big)\,\mathbf{U}\,\mathbf{U}^{-1} + h.c. \\
&= -\Big(\mathrm{D}^\alpha\,\mathrm{W}_\alpha + \mathbf{U}^{-1}\,(\mathrm{D}^\alpha\mathbf{U})\,\mathrm{W}_\alpha\Big)\,\mathbf{U}^{-1} + \mathrm{W}^\alpha\,\mathbf{U}^{-1}\Big(\mathrm{D}_\alpha\mathbf{U}\Big)\,\mathbf{U}^{-1} + h.c. \\
&= -\bigg(\mathrm{D}^\alpha\,\mathrm{W}_\alpha + \Big\{\mathbf{U}^{-1}\,(\mathrm{D}^\alpha\mathbf{U})\,,\,\mathrm{W}_\alpha\Big\}\bigg)\,\mathbf{U}^{-1} + h.c.
\end{aligned} \tag{17}$$

The terms inside the parenthesis define a new derivation in superspace

$$
\begin{aligned}
\mathscr{D}^{\alpha}\,\mathrm{W}_{\alpha} &:= \mathrm{D}^{\alpha}\,\mathrm{W}_{\alpha} + \Big\{\mathbf{U}^{-1}\,(\mathrm{D}^{\alpha}\mathbf{U})\,,\,\mathrm{W}_{\alpha}\Big\} \\
&= \Big\{\mathrm{D}^{\alpha} + \mathbf{U}^{-1}\,(\mathrm{D}^{\alpha}\mathbf{U})\,,\,\mathrm{W}_{\alpha}\Big\} \\
&= \Big\{\mathbf{U}^{-1}\,\mathrm{D}^{\alpha}\,\mathbf{U}\,,\,\mathrm{W}_{\alpha}\Big\} \\
&= \Big\{\,\mathscr{D}^{\alpha}\,,\,\mathrm{W}_{\alpha}\Big\}
\end{aligned}
\tag{18}
$$

where $\mathscr{D}^{\alpha}$ is the differential operator

$$
\mathscr{D}^{\alpha} \equiv \mathbf{U}^{-1}\,\mathrm{D}^{\alpha}\,\mathbf{U} = \mathrm{D}^{\alpha} + \mathbf{U}^{-1}\,(\mathrm{D}^{\alpha}\mathbf{U}) \tag{19}
$$

Of course, this is the familiar superspace covariant derivative in the chiral representation and it appears naturally through this process without us asking for an extended symmetry covariance. Nevertheless, its emergence reflects the existence of the expected geometric interpretation. Writing explicitly the *h.c.*[9] part of equation (17) we get:

$$
\mathscr{E}^{(\mathbf{U})} = -\big(\,\mathscr{D}^{\alpha}\,\mathrm{W}_{\alpha}\,\big)\,\mathbf{U}^{-1} \;-\; \mathbf{U}^{-1}\big(\,\bar{\mathfrak{D}}^{\dot{\alpha}}\,\bar{\mathbb{W}}_{\dot{\alpha}}\,\big) = 0 \tag{20}
$$

where

$$
\bar{\mathfrak{D}}^{\dot{\alpha}}\,\bar{\mathbb{W}}_{\dot{\alpha}} = \Big\{\,\bar{\mathfrak{D}}^{\dot{\alpha}}\,,\,\bar{\mathbb{W}}_{\dot{\alpha}}\Big\}\,, \tag{21a}
$$

$$
\bar{\mathbb{W}}_{\dot{\alpha}} := -\big(\,\mathrm{W}_{\alpha}\,\big)^{\dagger} = -\frac{4c}{g^{2}}\,\mathrm{D}^{2}\Big[\mathbf{U}\,(\bar{\mathrm{D}}_{\dot{\alpha}}\mathbf{U}^{-1})\Big]\,, \tag{21b}
$$

$$
\bar{\mathfrak{D}}^{\dot{\alpha}} := \big(\,\mathscr{D}^{\alpha}\,\big)^{\dagger} = \mathbf{U}\,\bar{\mathrm{D}}^{\dot{\alpha}}\,\mathbf{U}^{-1} = \bar{\mathrm{D}}^{\dot{\alpha}} + \mathbf{U}\,(\bar{\mathrm{D}}^{\dot{\alpha}}\mathbf{U}^{-1}) \tag{21c}
$$

and $\bar{\mathfrak{D}}_{\dot{\alpha}}$ is the geometric superspace covariant derivative in the anti-chiral representation. Moreover, equation (20) can be written in the following form:

$$
\begin{aligned}
0 = -\,\mathscr{E}^{(\mathbf{U})}\,\mathbf{U} &= \mathscr{D}^{\alpha}\,\mathrm{W}_{\alpha} + \mathbf{U}^{-1}\,(\bar{\mathfrak{D}}^{\dot{\alpha}}\,\bar{\mathbb{W}}_{\dot{\alpha}})\,\mathbf{U} \\
&= \mathscr{D}^{\alpha}\,\mathrm{W}_{\alpha} + \mathbf{U}^{-1}\Big\{\,\mathbf{U}\,\bar{\mathrm{D}}^{\dot{\alpha}}\,\mathbf{U}^{-1}\,,\,\bar{\mathbb{W}}_{\dot{\alpha}}\,\Big\}\mathbf{U} \\
&= \mathscr{D}^{\alpha}\,\mathrm{W}_{\alpha} + \,\bar{\mathrm{D}}^{\dot{\alpha}}\Big[\mathbf{U}^{-1}\,\bar{\mathbb{W}}_{\dot{\alpha}}\,\mathbf{U}\Big]
\end{aligned}
\tag{22}
$$

This equation provides the correct hermitian conjugation operation that we should adopt for elements of the chiral representation ($\mathbf{f}$)

$$
\mathbf{f} \rightarrow\; \overline{\;\mathbf{f}\;} := \mathbf{U}^{-1}\,\big(\,\mathbf{f}\,\big)^{\dagger}\,\mathbf{U} \tag{23}
$$

$$
\mathscr{D}_{\alpha} \rightarrow\; \bar{\mathscr{D}}_{\dot{\alpha}} = \bar{\mathrm{D}}_{\dot{\alpha}} \tag{23a}
$$

[9] We follow the conventions of *Superspace* [34].

$$\mathrm{W}_\alpha \to \bar{\mathrm{W}}_{\dot\alpha} = \mathbf{U}^{-1}\, \bar{\mathfrak{W}}_{\dot\alpha}\, \mathbf{U} \tag{23b}$$

Similarly, using (20) to evaluate $\mathbf{U}\,\mathscr{E}^{(\mathbf{U})}$ we derive the meaning of hermitian conjugation for the anti-chiral representation ($\mathfrak{f}$)

$$\mathfrak{f} \to \overline{\mathfrak{f}} := \mathbf{U}\left(\mathfrak{f}\right)^{\dagger}\mathbf{U}^{-1} \tag{24}$$

$$\bar{\mathfrak{D}}_{\dot\alpha} \to \mathfrak{D}_\alpha = \mathrm{D}_\alpha \tag{24a}$$

$$\bar{\mathfrak{W}}_{\dot\alpha} \to \mathfrak{W}_\alpha = \mathbf{U}\,\mathrm{W}_\alpha\,\mathbf{U}^{-1} \tag{24b}$$

In the conventional, geometric derivation of SYM these hermitian conjugation rules (—) are introduced as necessary modifications of the ordinary hermitian conjugation (†) to ensure that the pair of supersymmetric covariant derivatives $\bar\nabla_{\dot\alpha}\,,\ \nabla_\alpha = \overline{\left(\bar\nabla_{\dot\alpha}\right)}$ transform the same way under the group G action. This is in contrast with this approach, where the proper notion of hermitian conjugation unfolds from the consistency of the equations of motion (15) generated by action (13). It is also evident that the two representations are related by a similarity transformation $S\,:\ \mathfrak{f} \to \mathbf{f} = \mathbf{U}^{-1}\,\mathfrak{f}\,\mathbf{U}$ and the following diagram commutes:

$$\begin{array}{ccc} \mathfrak{f} & \xrightarrow{\ S\ } & \mathbf{f} = \mathbf{U}^{-1}\,\mathfrak{f}\,\mathbf{U} \\ {\scriptstyle(-)}\big\downarrow & & \big\downarrow{\scriptstyle(-)} \\ \overline{\mathfrak{f}} = \mathbf{U}\left(\mathfrak{f}\right)^{\dagger}\mathbf{U}^{-1} & \xrightarrow{\ S\ } & \overline{\mathbf{f}} = \left(\mathfrak{f}\right)^{\dagger} \end{array} \tag{24}$$

Using different similarity transformations we find other representations, like the vector representation which has the characteristic of restoring the ordinary hermitian conjugation.

Equation (22) can be further simplified, due to the identity

$$\mathscr{D}^\alpha\,\mathrm{W}_\alpha = \bar{\mathscr{D}}^{\dot\alpha}\,\bar{\mathrm{W}}_{\dot\alpha} \tag{25}$$

which reflects the hermiticity of $\mathscr{D}^\alpha\,\mathrm{W}_\alpha$ under rule (23) and can be verified in a straightforward manner (see Appendix A). In the geometric derivation, this identity is a Bianchi identity and follows from the tedious process of solving the Jacobi identities, once the set of conventional constraints[10] have been imposed (see [15] and [34, 35] for review). With that in mind, the on-shell equation of motion for superfield $\mathbf{U}$ is:

$$\mathscr{E} \equiv \mathscr{D}^\alpha\,\mathrm{W}_\alpha = 0 \tag{26}$$

[10] For SYM the set of conventional constraints are (1.) the supertorsion is equal to the flat superspace one, (2.) the spacetime covariant derivative is determined by the pair of spinorial covariant derivatives (SUSY algebra) and (3.) the existence of covariantly (anti-)chiral representations.

which is precisely the dynamical equation of SYM. Additionally, using solutions (16) we can integrate out superfields $\omega_\alpha$ and $\mathcal{W}_\alpha$

$$S = -\frac{2c}{g^2}\,\mathrm{Tr}\left[\int d^8z \left\{\mathbf{U}^{-1}(\mathrm{D}^\alpha\mathbf{U})\,\bar{\mathrm{D}}^2\left[\mathbf{U}^{-1}\,(\mathrm{D}_\alpha\mathbf{U})\right]\right\}\right] + c.c. \tag{27}$$

to find that action (13) is equivalent to the familiar SYM action, written in terms of **U**.

Expressing, equations (15) in terms of the **V** superfield and using (25), we find the following system of equations.

$$\omega_\alpha = -g\,\mathbf{V}\,\omega_\alpha - g\,\mathrm{D}_\alpha\mathbf{V}\,, \tag{28a}$$

$$\mathcal{W}_\alpha = -g\,\mathcal{W}_\alpha\,\mathbf{V} + \frac{4c}{g^2}\,\bar{\mathrm{D}}^2\omega_\alpha\,, \tag{28b}$$

$$\omega^\alpha\,\mathcal{W}_\alpha - \mathrm{D}^\alpha\mathcal{W}_\alpha = 0\,. \tag{28c}$$

Using (28a) and (28b) to calculate derivatives of $\omega_\alpha$ and $\mathcal{W}_\alpha$ and substituting into (28c) we find:

$$\mathcal{E} = \mathrm{D}^\gamma\bar{\mathrm{D}}^2\mathrm{D}_\gamma\mathbf{V} = -\frac{g}{4c}\,\mathbf{J} \tag{29}$$

where

$$\mathbf{J} = \omega^\alpha\,\mathcal{W}_\alpha + g\,\mathrm{D}^\alpha\left[\mathcal{W}_\alpha\,\mathbf{V}\right] + \frac{4c}{g}\,\mathrm{D}^\alpha\bar{\mathrm{D}}^2\left[\mathbf{V}\,\omega_\alpha\right]. \tag{30}$$

Equation (29) provides the desired augmentation of the corresponding free equation of motion (6) due to the presence of the source **J** which captures consistent self-interactions. This expression is non-perturbative and **J** satisfies the consistency conditions $\mathrm{D}^2\mathbf{J} = 0 = \bar{\mathrm{D}}^2\mathbf{J}$ as a result of the system of equations (28). Note, that only the first term ($\omega^\alpha\,\mathcal{W}_\alpha$) of **J** is required to appear in the action (10). As mentioned previously, its characteristic is that is algebraic, meaning it does not depend on derivatives of superfields and therefore does not require higher order corrections, thus allowing us to acquire the full, non-perturbative answer in a simple closed form. This highlights the value of the first-order description (1), without which writing such a term would not be possible. The rest of the terms in **J**, have a different origin. They correspond to the non-linearities in the relation between $\omega_\alpha$ ($\mathcal{W}_\alpha$) and **V** due to the presence of $\mathbf{U}^{-1}$ in (16).

Action (27) is, of course, invariant under an extended, non-abelian symmetry group **G** with respect to the following gauge transformation

$$\mathbf{U} \to \mathbf{U}' = \bar{\mathfrak{g}}\,\mathbf{U}\,\mathfrak{g} \tag{31}$$

were $\mathfrak{g}$ is a chiral element ($\bar{\mathrm{D}}_{\dot\alpha}\mathfrak{g} = 0$) of group **G**. The items $\omega_\alpha$, $\mathrm{W}_\alpha$, and $\mathscr{D}_\alpha$ as constructed above, under (31) acquire the proper transformations as superconnection, supercurvature and super covariant derivative respectively. It is important to highlight that neither the gauge invariance of the action nor the specific gauge transformation properties of the aforementioned elements were prerequisites in the beginning and were not taken into account in formulating (10) or (13). This is an emergent property of the theory originating from the consistency of the interactions. To see that in detail, we can rewrite (10) in the following way:

$$S = \int d^8z \left\{ \mathcal{W}^{I\alpha}\, \Omega^J_\alpha\, g_{IJ} + \mathcal{W}^{I\alpha}\, \mathrm{D}_\alpha \mathrm{V}^J\, g_{IJ} - 2c\, \Omega^{I\alpha}\, \bar{\mathrm{D}}^2 \Omega^J_\alpha\, g_{IJ} + \mathrm{g}\, \mathcal{W}^{I\alpha}\, \mathrm{V}^J\, \Omega^K_\alpha\, f_{IJK} \right\} + c.c. \tag{32}$$

where we expanded each matrix in a randomly chosen complete basis and we performed the trace. These operations are captured by two arrays of numbers (a.) the symmetric $g_{IJ}$ which corresponds to the trace of the product of two elements of the basis and (b.) $f_{IJK}$ which corresponds to the trace of the product of three elements of the basis. Now we can check again the requirements for the cubic term to be a consistent and non-trivial interaction. The cubic interaction is of the form $\mathscr{L}_{\text{int}} \propto \mathrm{V}^K \mathrm{J}_K$ and in order for it to be consistent (a.) the current $\mathrm{J}_K = \mathcal{W}^{I\alpha}\, \Omega^J_\alpha f_{IJK} + c.c.$ must satisfy the (on-shell) conservation equations $\mathrm{D}^2 \mathrm{J}_K = 0$ and (b.) $\mathscr{L}_{\text{int}}$ must be non-trivial, hence it must not vanish on-shell:

$$\mathrm{D}^2\, \mathrm{J}_K \Big|_{\delta S_o = 0} = 0 \Longrightarrow f_{IJK} + f_{JIK} = 0\,, \tag{33}$$

$$\mathscr{L}_{\text{int}} \Big|_{\delta S_o = 0} \neq 0 \Longrightarrow f_{IJK} + f_{IKJ} = 0\,,\; f_{IJK} \neq 0\,. \tag{34}$$

Therefore, the consistency of the cubic interaction requires the existence of a non-trivially zero array of numbers $f_{IJK}$ which is completely antisymmetric

$$f_{IJK} = -f_{JIK} = -f_{IKJ}\,. \tag{35}$$

Thus, $f_{IJK}$ can be understood as the structure constants of some Lie algebra and $g_{IJ}$ corresponds to the Cartan-Killing form.

Finally, we consider the coupling of matter to this theory. The presence of matter will contribute additional conserved supercurrents that will be added as sources to $\mathbf{J}$ in (29). For the case of chiral superfields, the conserved supercurrent is $\mathbf{J} = \boldsymbol{\Phi}\, \bar{\boldsymbol{\Phi}}$[11]. This current generates the following cubic interaction

$$S_1 = \mathrm{Tr}\left[ \int d^8z \left\{ \mathrm{g}\, \bar{\boldsymbol{\Phi}}\, \mathbf{V}\, \boldsymbol{\Phi} \right\} \right] \tag{36}$$

which when added to the free theory of massless chiral supermultiplets gives

$$S = \mathrm{Tr}\left[ \int d^8z \left\{ \bar{\boldsymbol{\Phi}}\, \boldsymbol{\Phi} + \mathrm{g}\, \bar{\boldsymbol{\Phi}}\, \mathbf{V}\, \boldsymbol{\Phi} \right\} \right] = \mathrm{Tr}\left[ \int d^8z \left\{ \bar{\boldsymbol{\Phi}} \left( \mathbb{1} + \mathrm{g}\mathbf{V} \right) \boldsymbol{\Phi} \right\} \right]$$

[11] It is easy to see that $\mathrm{D}^2(\boldsymbol{\Phi}\, \bar{\boldsymbol{\Phi}})$ and $\mathrm{D}^2(\boldsymbol{\Phi}\, \bar{\boldsymbol{\Phi}})$ vanish on-shell and the corresponding interaction is non-trivial.

$$= \mathrm{Tr}\left[\int d^8z \left\{\bar{\mathbf{\Phi}}\,\mathbf{U}\,\mathbf{\Phi}\right\}\right]. \tag{37}$$

This is precisely, the usual coupling of matter to SYM, expressed in terms of $\mathbf{U}$. It is derived by adding a single, consistent, non-trivial cubic interaction term (36). At this point, the geometrical interpretation of SYM is complete. All matter must now behave as elements of an effective associated vector bundle with bridge $\mathbf{U} = \mathbb{1} + g\,\mathbf{V}$.

As emphasized previously, the derivation of the superspace description of SYM and its coupling to matter presented here (α.) does not rely on the assumption of an extended group of symmetries known in advance and (β.) the simple polynomial form of the action, as opposed to an infinite series of self-interactions, is achieved without the need to introduce a special gauge. These characteristics may raise concerns regarding the extraction of the spacetime component Lagrangian from the superspace action. Common practices for this process in SYM involve fixing a gauge (WZ) to convert all interactions into the standard QFT form, which may lead one to wonder how to achieve this without utilizing the (*emergent*) symmetry properties of the theory. However, there is no cause for concern because the process of extracting the spacetime Lagrangian from the superspace action does not require selecting a specific gauge. This is due to the simple fact that the SYM superspace action does not depend on all the components of all superfields involved, but only on a specific subset. As mentioned, a convenient and widely used technique to identify this subset of component fields is to $\theta$-expand all superfields and choose an appropriate gauge (WZ) which eliminates extraneous components and determines the collection of spacetime fields appearing in the component Lagrangian. A different methodology is the *component projection* method which performs the Berezin integration of the superspace action along the $\theta$ directions by evaluating the superspace derivatives of the Lagrangian at $\theta = 0$. Then using the algebra of these derivatives we collect terms into familiar forms. These provide a natural definition of the spacetime components of the theory without requiring a specific gauge. Detailed examples of this methodology for gauge theories of arbitrary spin can be found at [36]. To conclude, we present the process of reducing the superspace theories (13) and (37) to their component forms.

Action (13) can be written in the following form

$$S = \mathrm{Tr}\left[\int d^8z \left\{\mathrm{W}^{\alpha}\,\omega_{\alpha} + \mathrm{W}^{\alpha}\,\Xi_{\alpha} - \frac{2c}{g^2}\,\omega^{\alpha}\,\bar{\mathrm{D}}^2\omega_{\alpha}\right\}\right] + c.c. \tag{38}$$

where $\mathrm{W}_{\alpha} = \mathscr{W}_{\alpha}\,\mathbf{U}$ and $\Xi_{\alpha}$ is a Lie-algebra valued 'Maurer-Cartan' spinor

$$\Xi_{\alpha} = \mathbf{U}^{-1}\,(\mathrm{D}_{\alpha}\mathbf{U}) \tag{39}$$

which satisfies the following Maurer-Cartan like equation:

$$\mathrm{D}_{(\beta}\Xi_{\alpha)} + \Xi_{(\beta}\,\Xi_{\alpha)} = 0 \tag{40}$$

This form of the action is especially useful for deriving the spacetime components of the theory as it allows an immediate comparison with the free limit of (10) —superymmetric Maxwell theory (1)— by replacing $\Xi_\alpha$ with $\mathrm{D}_\alpha \mathbf{V}$ and $\frac{c}{g^2}$ with $c$. Up to total derivatives, the spacetime Lagrangian under the trace sign corresponding to (38) is

$$\mathscr{L} = \bar{\mathrm{D}}^2 \mathrm{D}^2 \Big\{ \mathrm{W}^\alpha\, \omega_\alpha + \mathrm{W}^\alpha\, \Xi_\alpha - \frac{2c}{g^2}\, \omega^\alpha\, \bar{\mathrm{D}}^2 \omega_\alpha \Big\}\Big|_{\theta=0,\bar{\theta}=0} + h.c. \tag{41}$$

where the meaning of hermitian conjugation operation can include similarity transformations due to $Tr[A^\dagger] = Tr[S A^\dagger S^{-1}]$. This is particularly useful since we can employ the chiral representation hermitian conjugation (23). To demonstrate the process of extracting the corresponding spacetime Lagrangian we focus only on the bosonic part of this Lagrangian, by ignoring all the terms that include fermions: $\mathscr{L}^{(B)} = \mathscr{L}\Big|_{\text{Bosons}}$. By distributing the derivatives in (41), using their algebra and collecting terms we find:

$$\begin{aligned}
\mathscr{L}^{(B)} = & -\frac{1}{2}\{\mathrm{D}^2, \bar{\mathrm{D}}^{\dot\alpha}\}\mathrm{W}^\alpha\Big| \left( \bar{\mathrm{D}}_{\dot\alpha}\, \omega_\alpha\Big| + \bar{\mathrm{D}}_{\dot\alpha}\, \Xi_\alpha\Big| \right) \\
& -\frac{1}{2}\, \bar{\mathrm{D}}^{\dot\alpha}\, \mathrm{W}^\alpha\Big| \left( \{\mathrm{D}^2, \bar{\mathrm{D}}_{\dot\alpha}\}\, \omega_\alpha\Big| + \{\mathrm{D}^2, \bar{\mathrm{D}}_{\dot\alpha}\}\, \Xi_\alpha\Big| \right) \\
& -\frac{1}{8}\{\bar{\mathrm{D}}^2, \mathrm{D}^{(\beta}\}\mathrm{W}^{\alpha)}\Big| \left( \mathrm{D}_{(\beta}\, \omega_{\alpha)}\Big| + \mathrm{D}_{(\beta}\, \Xi_{\alpha)}\Big| \right) \\
& -\frac{1}{4}\{\bar{\mathrm{D}}^2, \mathrm{D}^\alpha\}\mathrm{W}_\alpha\Big| \left( \mathrm{D}^\beta\, \omega_\beta\Big| + \mathrm{D}^\beta\, \Xi_\beta\Big| \right) \\
& -\frac{1}{8}\, \mathrm{D}^{(\beta}\, \mathrm{W}^{\alpha)}\Big|\ \{\bar{\mathrm{D}}^2, \mathrm{D}_{(\beta}\}\, \Xi_{\alpha)}\Big| \\
& -\frac{1}{4}\, \mathrm{D}^\alpha\, \mathrm{W}_\alpha\Big|\ \{\bar{\mathrm{D}}^2, \mathrm{D}^\beta\}\, \Xi_\beta\Big| \\
& +\frac{c}{8g^2}\{\bar{\mathrm{D}}^2, \mathrm{D}^{(\beta}\}\, \omega^{\alpha)}\Big| \left( \{\bar{\mathrm{D}}^2, \mathrm{D}_{(\beta}\}\, \omega_{\alpha)}\Big| + 2i\, \partial_{(\beta}{}^{\dot\alpha}\, \bar{\mathrm{D}}_{\dot\alpha}\, \omega_{\alpha)}\Big| - \frac{g^2}{c}\, \mathrm{D}_{(\beta}\, \mathrm{W}_{\alpha)}\Big| \right) \\
& +\frac{c}{4g^2}\{\bar{\mathrm{D}}^2, \mathrm{D}^\alpha\}\, \omega_\alpha\Big| \left( \{\bar{\mathrm{D}}^2, \mathrm{D}^\beta\}\, \omega_\beta\Big| + 2i\, \partial^{\beta\dot\beta}\, \bar{\mathrm{D}}_{\dot\beta}\, \omega_\beta\Big| - \frac{g^2}{c}\, \mathrm{D}^\beta\, \mathrm{W}_\beta\Big| \right) \\
& +\frac{c}{2g^2}\, \bar{\mathrm{D}}^{\dot\alpha}\, \omega^\alpha\Big|\ \Box\ \bar{\mathrm{D}}_{\dot\alpha}\, \omega_\alpha\Big| \\
& + h.c.
\end{aligned} \tag{42}$$

By integrating out the following auxiliary components $\{\mathrm{D}^2, \bar{\mathrm{D}}^{\dot\alpha}\}\mathrm{W}^\alpha\Big|$, $\bar{\mathrm{D}}^{\dot\alpha}\, \mathrm{W}^\alpha\Big|$, $\{\bar{\mathrm{D}}^2, \mathrm{D}^{(\beta}\}\mathrm{W}^{\alpha)}\Big|$, $\{\bar{\mathrm{D}}^2, \mathrm{D}^\alpha\}\mathrm{W}_\alpha\Big|$, $\{\bar{\mathrm{D}}^2, \mathrm{D}^{(\beta}\}\, \omega^{\alpha)}\Big|$ and $\{\bar{\mathrm{D}}^2, \mathrm{D}^\alpha\}\, \omega_\alpha\Big|$ we get the following equations:

$$\{\mathrm{D}^2, \bar{\mathrm{D}}^{\dot\alpha}\}\mathrm{W}^\alpha\Big| \quad : \qquad \bar{\mathrm{D}}_{\dot\alpha}\, \omega_\alpha\Big| + \bar{\mathrm{D}}_{\dot\alpha}\, \Xi_\alpha\Big| = 0 \tag{43a}$$

$$\bar{\mathrm{D}}^{\dot\alpha}\,\mathrm{W}^{\alpha}\Big| \; : \qquad \left\{\mathrm{D}^2,\bar{\mathrm{D}}_{\dot\alpha}\right\}\omega_\alpha\Big| \;+\; \left\{\mathrm{D}^2,\bar{\mathrm{D}}_{\dot\alpha}\right\}\Xi_\alpha\Big| = 0 \tag{43b}$$

$$\left\{\bar{\mathrm{D}}^2,\mathrm{D}^{(\beta}\right\}\mathrm{W}^{\alpha)}\Big| \; : \qquad \mathrm{D}_{(\beta}\,\omega_{\alpha)}\Big| \;+\; \mathrm{D}_{(\beta}\,\Xi_{\alpha)}\Big| = 0 \tag{43c}$$

$$\left\{\bar{\mathrm{D}}^2,\mathrm{D}^{\alpha}\right\}\mathrm{W}_{\alpha}\Big| \; : \qquad \mathrm{D}^{\beta}\,\omega_{\beta}\Big| \;+\; \mathrm{D}^{\beta}\,\Xi_{\beta}\Big| = 0 \tag{43d}$$

$$\left\{\bar{\mathrm{D}}^2,\mathrm{D}^{(\beta}\right\}\omega^{\alpha)}\Big| \; : \; 2\left\{\bar{\mathrm{D}}^2,\mathrm{D}_{(\beta}\right\}\omega_{\alpha)}\Big| \;+\; 2i\,\partial_{(\beta}{}^{\dot\alpha}\,\bar{\mathrm{D}}_{\dot\alpha}\,\omega_{\alpha)}\Big| \;-\; \frac{g^2}{c}\,\mathrm{D}_{(\beta}\,\mathrm{W}_{\alpha)}\Big| = 0 \tag{43e}$$

$$\left\{\bar{\mathrm{D}}^2,\mathrm{D}^{\alpha}\right\}\omega_{\alpha}\Big| \; : \qquad 2\left\{\bar{\mathrm{D}}^2,\mathrm{D}^{\beta}\right\}\omega_{\beta}\Big| \;+\; 2i\,\partial^{\beta\dot\beta}\,\bar{\mathrm{D}}_{\dot\beta}\,\omega_{\beta}\Big| \;-\; \frac{g^2}{c}\,\mathrm{D}^{\beta}\,\mathrm{W}_{\beta}\Big| = 0 \tag{43f}$$

Substituting back to $\mathscr{L}^{(B)}$ we find

$$\begin{aligned}\mathscr{L}^{(B)} = &-\frac{g^2}{32c}\,\mathbf{F}^{\alpha\beta}\,\mathbf{F}_{\alpha\beta} \;-\frac{1}{4}\,\mathbf{F}^{\alpha\beta}\left(\mathrm{D}_{(\beta}\bar{\mathrm{D}}^2\,\Xi_{\alpha)}\Big|\right)\\ &-\frac{g^2}{16c}\,\mathbf{d}^2 - \frac{1}{2}\,\mathbf{d}\left(\mathrm{D}^{\beta}\bar{\mathrm{D}}^2\,\Xi_{\beta}\Big|\right)\\ &+h.c.\end{aligned} \tag{44}$$

where $\mathbf{F}_{\alpha\beta} := \mathrm{D}_{(\beta}\,\mathrm{W}_{\alpha)}\Big|$ and $\mathbf{d} := \mathrm{D}^{\alpha}\,\mathrm{W}_{\alpha}\Big|$.

For the free theory ($\Xi_\alpha \to \mathrm{D}_\alpha\mathbf{V}$, $\frac{c}{g^2} \to c$) we get:

$$\mathrm{D}_{(\beta}\bar{\mathrm{D}}^2\,\Xi_{\alpha)}\Big| = \mathrm{D}_{(\beta}\bar{\mathrm{D}}^2\mathrm{D}_{\alpha)}\mathbf{V}\Big| = i\,\partial_{(\beta}{}^{\dot\alpha}\,\bar{\mathrm{D}}_{\dot\alpha}\mathrm{D}_{\alpha)}\mathbf{V}\Big| = \frac{i}{2}\,\partial_{(\beta}{}^{\dot\alpha}\,[\bar{\mathrm{D}}_{\dot\alpha},\mathrm{D}_{\alpha)}]\mathbf{V}\Big| = i\,\partial_{(\beta}{}^{\dot\alpha}\,\mathbf{A}_{\alpha)\dot\alpha}\;, \tag{45a}$$

$$\mathrm{D}^{\alpha}\bar{\mathrm{D}}^2\Xi_{\alpha}\Big| = \mathrm{D}^{\alpha}\bar{\mathrm{D}}^2\mathrm{D}_{\alpha}\,\mathbf{V}\Big| = \mathbf{D} \tag{45b}$$

where the fields $\mathbf{A}_{\alpha\dot\alpha} := \frac{1}{2}\,[\bar{\mathrm{D}}_{\dot\alpha},\mathrm{D}_{\alpha)}]\mathbf{V}\Big|$ are real, gauge bosons and they are defined up to a total derivative $\mathbf{A}_{\alpha\dot\alpha} \sim \mathbf{A}_{\alpha\dot\alpha} + \partial_{\alpha\dot\alpha}\mathbf{\Lambda}$ due to the identity $\partial_{(b}{}^{\dot\alpha}\partial_{\alpha)\dot\alpha} = 0$. The fields $\mathbf{D} := \mathrm{D}^{\gamma}\bar{\mathrm{D}}^2\mathrm{D}_{\gamma}\,\mathbf{V}\Big|$ are also real and are the auxiliary supersymmetric components required by off-shell supersymmetry. In this case, equation (44) provides the first order description of the bosonic part of supersymmetric Maxwell theory. Note that this method of deriving the component Lagrangian eliminates the need to choose a special (WZ) gauge and does not require prior knowledge of the gauge properties of the components. The bosonic Lagrangian depends solely on two real fields: (1) the equivalence class $[\mathbf{A}_{\alpha\dot\alpha}]$ and (2) $\mathbf{D}$. The corresponding superfield that captures this (off-shell) supersymmetric multiplet is an equivalence class of superfields $[\mathbf{V}]$ with representative[12]:

$$\mathbf{V} = \theta^{\alpha}\,\bar\theta^{\dot\alpha}\,\mathbf{A}_{\alpha\dot\alpha} + \bar\theta^2\,\theta^{\alpha}\,\boldsymbol{\psi}_{\alpha} + \theta^2\,\bar\theta^{\dot\alpha}\bar{\boldsymbol{\psi}}_{\dot\alpha} + \frac{1}{2}\,\theta^2\,\bar\theta^2\,\mathbf{D}\;. \tag{46}$$

[12]As demonstrated in [37], supersymmetric transformations do not preserve the structure of this representative superfield but remain within its equivalence class.

This representative precisely aligns with the WZ gauge of a generic superfield **V**, however it is not obtained through exploring the gauge transformation properties of its components but rather by examining the component dependence of the Lagrangian.

For the interacting theory ( $\mathrm{D}_{(\beta}\Xi_{\alpha)} = -\Xi_{(\beta}\,\Xi_{\alpha)}$ ) we get:

$$\begin{aligned}\mathrm{D}_{(\beta}\bar{\mathrm{D}}^2\,\Xi_{\alpha)}\Big| &= [\mathrm{D}_{(\beta},\bar{\mathrm{D}}^2]\,\Xi_{\alpha)}\Big| + \bar{\mathrm{D}}^2\mathrm{D}_{(\beta}\,\Xi_{\alpha)}\Big| = i\,\partial_{(\beta}{}^{\dot\alpha}\,\bar{\mathrm{D}}_{\dot\alpha}\Xi_{\alpha)}\Big| + \bar{\mathrm{D}}^{\dot\alpha}\Xi_{(\beta}\Big|\bar{\mathrm{D}}_{\dot\alpha}\Xi_{\alpha)}\Big| \\ &= C^{\dot\alpha\dot\beta}\left(i\,\partial_{(\beta\dot\beta}\,\bar{\mathrm{D}}_{\dot\alpha}\Xi_{\alpha)}\Big| + \frac{1}{2}\left[\,\bar{\mathrm{D}}_{\dot\beta}\Xi_{(\beta}\Big|\,,\,\bar{\mathrm{D}}_{\dot\alpha}\Xi_{\alpha)}\Big|\,\right]\right)\end{aligned} \tag{47}$$

therefore the bosonic part of the Lagrangian depends on two component fields $\bar{\mathrm{D}}_{\dot\alpha}\Xi_{\alpha}\Big|$ and $\mathrm{D}^{\alpha}\bar{\mathrm{D}}^2\Xi_{\alpha}\Big|$. The first one is the dynamical gauge boson

$$\bar{\mathrm{D}}_{\dot\alpha}\Xi_{\alpha}\Big| = \frac{1}{2}\,\mathbf{U}^{-1}\,[\bar{\mathrm{D}}_{\dot\alpha},\mathrm{D}_{\alpha}]\mathbf{U}\Big| + \frac{i}{2}\,\mathbf{U}^{-1}\,\partial_{\alpha\dot\alpha}\mathbf{U}\Big| \overset{(12)\,\&\,(46)}{=} g\,\mathbf{A}_{\alpha\dot\alpha} \tag{48}$$

and the second one is the auxiliary boson $g\,\mathbf{D}$. Hence, the bosonic part of the Lagrangian (44) takes the form:

$$\begin{aligned}\mathscr{L}^{(B)} = &-\frac{g^2}{32c}\,\mathbf{F}^{\alpha\beta}\,\mathbf{F}_{\alpha\beta} - i\,\frac{g}{4}\,\mathbf{F}^{\alpha\beta}\,C^{\dot\alpha\dot\beta}\,\boldsymbol{f}_{\alpha\beta\dot\alpha\dot\beta} + h.c. \\ &-\frac{g^2}{16c}\,\mathbf{d}^2 - \frac{g}{2}\,\mathbf{d}\,\mathbf{D} + h.c.\end{aligned} \tag{49}$$

where $\boldsymbol{f}_{\alpha\beta\dot\alpha\dot\beta}$ is the YM field strength

$$\boldsymbol{f}_{\alpha\beta\dot\alpha\dot\beta} = \partial_{(\beta\dot\beta}\,\mathbf{A}_{\alpha)\dot\alpha} - i\,\frac{g}{2}\left[\,\mathbf{A}_{(\beta\dot\beta}\,,\,\mathbf{A}_{\alpha)\dot\alpha}\,\right] \tag{50}$$

This is the first order description of YM. Similarly, the component projection of action (37) includes the following matter fermionic spacetime Lagrangian:

$$\mathscr{L}^{(F)}_{\text{matter}} = \bar\lambda^{\dot\alpha}\left\{i\,\partial^{\alpha}{}_{\dot\alpha}\,\lambda_{\alpha} + g\,\mathbf{A}^{\alpha}{}_{\dot\alpha}\,\lambda_{\alpha}\right\} + h.c. \tag{51}$$

where $\mathrm{D}_{\alpha}\Phi\Big| = \lambda_{\alpha}$.

# 3 Summary and Outlook

The construction of low spin, interacting, supersymmetric gauge theories is based on arguments rooted on the geometric interpretation and characteristics of these theories as well as the behavior of matter within such geometric frameworks. In both Super Yang-Mills and Supergravity, we rely on these intrinsic geometrical properties to postulate the existence of supersymmetric covariant derivatives. We then proceed to calculate their algebra and impose various (*conventional*) constraints to minimize the number of independent fields and reduce the degrees of freedom. This involves the cumbersome procedure of solving these constraints and their associated integrability conditions (such as Jacobi identities) by introducing unconstrained prepotential superfields which are used to write covariant geometric quantities and invariant actions. This achievement comes at the cost of non-polynomial actions with infinite non-linearities, which significantly complicates the study of non-perturbative effects. Moreover, this geometric approach is not suitable for the construction of interacting supersymmetric gauge theories involving higher spins, as these theories currently lack a geometrical understanding.

It is desirable to find an alternative approach to the foundations of interacting supersymmetric gauge theories, one which is not based on geometry, but instead focuses on the field theoretic consistency requirements of self-interactions that extend the initial free theory. Such an approach, using familiar arguments, may seem à priori not preferable due to its perturbative nature that generates iteratively infinite terms and pose challenges for their resummation to a closed-form action. Nevertheless, as has been demonstrated previously, the use of a first-order description has the capacity to sidestep these issues and furnish a closed form polynomial action.

In this work, we present a non-geometric derivation of $\mathcal{N} = 1$ Super Yang-Mills as the consistent self-interacting theory that builds upon the initial free vector supermultiplet, by utilizing the superspace first-order description developed in [26, 27]. Its major benefits, and differences from other derivations, include: (1.) the action is in polynomial form and features a single cubic interaction term, (2.) the invariance of the theory under the extended, non-abelian, symmetry group emerges as a result of the procedure rather than being initially required, (3.) the geometrical interpretation of the theory arises through the appearance of an effective variable $\mathbf{U} = \mathbb{1} + g\,\mathbf{V}$, which renders the initial background unperceivable, forces all matter to be elements of an effective associated G-bundle and transforms ordinary superspace derivatives into covariant derivatives of a principal G-bundle.

The existence and effectiveness of this approach in constructing interacting supersymmetric theories non-perturbatively, without prior assumptions of their geometric and symmetry structure, unveils several intriguing possibilities. Firstly, it offers an alternative route to understand the quantum properties of these theories, as path integral quantization becomes simpler due to the polynomial nature of the action. Secondly, it enables the study of systems lacking a geometric understanding of their dynamics, such as higher spins. An exciting task is to investigate whether Supergravity can be described in this manner. It is important to note that the non-geometrical interpretation of SYM presented here does not replace the geometrical understanding of low-spin gauge theories; rather, it highlights and pays tribute to the broad scope of these theories by offering insights into the origin of such structures.

## Acknowledgments

This work is supported by the endowment of the Clark Leadership Chair in Science at the University of Maryland, College Park. The author expresses gratitude to Prof. S. James Gates Jr. for engaging discussions and gratefully acknowledges the hospitality of the Physics Department at the University of Maryland, College Park.

## Appendix A

In this appendix we offer a proof of the hermiticity property (25)

$$\mathscr{D}^{\alpha}\,\mathrm{W}_{\alpha} = \bar{\mathscr{D}}^{\dot{\alpha}}\,\bar{\mathrm{W}}_{\dot{\alpha}} \tag{A.1}$$

The left hand side of the equation, after some straightforward algebra, can be written as follows

$$\begin{aligned}\frac{g^2}{4c}\,\mathscr{D}^{\alpha}\,\mathrm{W}_{\alpha} = {}& \mathrm{D}^{\alpha}\bar{\mathrm{D}}^2\big[\mathbf{U}^{-1}\left(\mathrm{D}_{\alpha}\mathbf{U}\right)\big] + \Big\{\mathbf{U}^{-1}\left(\mathrm{D}^{\alpha}\mathbf{U}\right),\ \bar{\mathrm{D}}^2\big[\mathbf{U}^{-1}\left(\mathrm{D}_{\alpha}\mathbf{U}\right)\big]\Big\}\\ = {}& -\mathrm{D}^{\gamma}\bar{\mathrm{D}}^2\mathrm{D}_{\gamma}\mathbf{U}^{-1}\,\mathbf{U} - \bar{\mathrm{D}}^{\dot{\alpha}}\mathrm{D}^{\alpha}\mathbf{U}^{-1}\,\mathrm{D}_{\alpha}\bar{\mathrm{D}}_{\dot{\alpha}}\mathbf{U} - 2\,\mathrm{D}^2\mathbf{U}^{-1}\,\bar{\mathrm{D}}^2\mathbf{U}\\ & - \bar{\mathrm{D}}^{\dot{\alpha}}\mathrm{D}^2\mathbf{U}^{-1}\,\bar{\mathrm{D}}_{\dot{\alpha}}\mathbf{U} - \mathrm{D}^2\bar{\mathrm{D}}^{\dot{\alpha}}\mathbf{U}^{-1}\,\bar{\mathrm{D}}_{\dot{\alpha}}\mathbf{U}\\ & - \mathrm{D}^{\alpha}\mathbf{U}^{-1}\,\mathrm{D}_{\alpha}\bar{\mathrm{D}}^2\mathbf{U} - \mathrm{D}^{\alpha}\mathbf{U}^{-1}\,\bar{\mathrm{D}}^2\mathrm{D}_{\alpha}\mathbf{U}\\ & - \bar{\mathrm{D}}^{\dot{\alpha}}\big[\mathrm{D}_{\alpha}\mathbf{U}^{-1}\,\mathbf{U}\,\bar{\mathrm{D}}_{\dot{\alpha}}\mathbf{U}^{-1}\,\mathrm{D}^{\alpha}\mathbf{U}\big]\,.\end{aligned} \tag{A.2}$$

The right hand side of the equation is

$$\begin{aligned}-\frac{g^2}{4c}\,\bar{\mathscr{D}}^{\dot{\alpha}}\,\bar{\mathrm{W}}_{\dot{\alpha}} = {}& \bar{\mathrm{D}}^{\dot{\alpha}}\Big[\mathbf{U}^{-1}\,\mathrm{D}^2\big[\mathbf{U}\left(\bar{\mathrm{D}}_{\dot{\alpha}}\mathbf{U}^{-1}\right)\big]\,\mathbf{U}\Big]\\ = {}& \mathrm{D}^{\gamma}\bar{\mathrm{D}}^2\mathrm{D}_{\gamma}\mathbf{U}^{-1}\,\mathbf{U} + \bar{\mathrm{D}}^{\dot{\alpha}}\mathrm{D}^{\alpha}\mathbf{U}^{-1}\,\mathrm{D}_{\alpha}\bar{\mathrm{D}}_{\dot{\alpha}}\mathbf{U} + 2\,\mathrm{D}^2\mathbf{U}^{-1}\,\bar{\mathrm{D}}^2\mathbf{U}\\ & + \bar{\mathrm{D}}^{\dot{\alpha}}\mathrm{D}^2\mathbf{U}^{-1}\,\bar{\mathrm{D}}_{\dot{\alpha}}\mathbf{U} + \mathrm{D}^2\bar{\mathrm{D}}^{\dot{\alpha}}\mathbf{U}^{-1}\,\bar{\mathrm{D}}_{\dot{\alpha}}\mathbf{U}\\ & + \mathrm{D}^{\alpha}\mathbf{U}^{-1}\,\mathrm{D}_{\alpha}\bar{\mathrm{D}}^2\mathbf{U} + \mathrm{D}^{\alpha}\mathbf{U}^{-1}\,\bar{\mathrm{D}}^2\mathrm{D}_{\alpha}\mathbf{U}\\ & + \bar{\mathrm{D}}^{\dot{\alpha}}\big[\mathrm{D}_{\alpha}\mathbf{U}^{-1}\,\mathbf{U}\,\bar{\mathrm{D}}_{\dot{\alpha}}\mathbf{U}^{-1}\,\mathrm{D}^{\alpha}\mathbf{U}\big]\,.\end{aligned} \tag{A.3}$$

Adding (A.2) and (A.3) gives (A.1).

## References

[1] X. Bekaert, "Geometric tool kit for higher-spin gravity (Part I): An introduction to the geometry of differential operators," Int. J. Mod. Phys. A **38**, 2330003 (2023), arXiv:2301.08069 [hep-th] (cit. on p. 2).

[2] X. Bekaert, "Geometric tool kit for higher spin gravity (Part II): An introduction to Lie algebroids and their enveloping algebras," Int. J. Mod. Phys. A **38**, 2330013 (2023), arXiv:2308.00724 [hep-th] (cit. on p. 2).

[3] R. H. Kraichnan, "Special-Relativistic Derivation of Generally Covariant Gravitation Theory," Phys. Rev. **98**, 1118–1122 (1955) (cit. on p. 2).

[4] A. Papapetrou, "Einstein's theory of gravitation and flat space," Proc. Roy. Irish Acad. A **52**, 11–23 (1948) (cit. on p. 2).

[5] S. N. Gupta, "Quantization of Einstein's gravitational field: general treatment," Proc. Phys. Soc. A **65**, 608–619 (1952) (cit. on p. 2).

[6] R. P. Feynman, *Conference on the role of gravitation in physics at the University of North Carolina, Chapel Hill [January 18–23, 1957, under the sponsorship of the International Union of Pure and Applied Physics, and others*, edited by C. DeWitt-Morette and D. Rickles, 1957 (cit. on p. 2).

[7] W. Thirring, "Lorentz-invariante gravitationstheorien," Fortschritte der Physik **7**, 79–101 (1959) (cit. on p. 2).

[8] V. I. Ogievetsky and I. V. Polubarinov, "Interacting field of spin 2 and the Einstein equations," Annals Phys. **35**, 167–208 (1965) (cit. on p. 2).

[9] V. I. Ogievetsky and I. V. Polubarinov, "On the meaning of gauge invariance," Nuovo Cim. **23**, 173–180 (1962) (cit. on p. 2).

[10] V. I. Ogievetsky and I. V. Polubarinov, "On interacting fields with definite spin," Zh. Eksp. Teor. Fiz. **45**, 237–245 (1963) (cit. on p. 2).

[11] V. I. Ogievetskii and I. V. Polubarinov, "Minimal Interactions between spin 0, 1/2, and 1 fields," Zh. Eksperim. i Teor. Fiz. **46** (1964) (cit. on p. 2).

[12] V. Ogiyevetsky and I. Polubarinov, "Theories of interacting fields with spin 1," Nuclear Physics **76**, 677–683 (1966) (cit. on p. 2).

[13] E. A. Ivanov, "Gauge Fields, Nonlinear Realizations, Supersymmetry," Phys. Part. Nucl. **47**, 508–539 (2016), arXiv:1604.01379 [hep-th] (cit. on p. 2).

[14] S. Deser, "Selfinteraction and gauge invariance," Gen. Rel. Grav. **1**, 9–18 (1970), arXiv:gr-qc/0411023 (cit. on p. 3).

[15] R. Grimm, M. Sohnius, and J. Wess, "Extended Supersymmetry and Gauge Theories," Nucl. Phys. B **133**, 275–284 (1978) (cit. on pp. 3, 9).

[16] J. Wess and B. Zumino, "Superspace Formulation of Supergravity," Phys. Lett. B **66**, 361–364 (1977) (cit. on p. 3).

[17] S. J. Gates Jr. and W. Siegel, "Understanding Constraints in Superspace Formulations of Supergravity," Nucl. Phys. B **163**, 519–545 (1980) (cit. on p. 3).

[18] S. Ferrara and B. Zumino, "Supergauge Invariant Yang-Mills Theories," Nucl. Phys. B **79**, 413 (1974) (cit. on p. 3).

[19] W. Siegel, "The Superfield Supergravity Action," (1977) (cit. on p. 3).

[20] W. Siegel and S. J. Gates Jr., "Superfield Supergravity," Nucl. Phys. B **147**, 77–104 (1979) (cit. on p. 3).

[21] I. Buchbinder, S. J. Gates, and K. Koutrolikos, "Higher Spin Superfield interactions with the Chiral Supermultiplet: Conserved Supercurrents and Cubic Vertices," Universe **4**, 6 (2018), arXiv:1708.06262 [hep-th] (cit. on p. 3).

[22] K. Koutrolikos, P. Kočí, and R. von Unge, "Higher Spin Superfield interactions with Complex linear Supermultiplet: Conserved Supercurrents and Cubic Vertices," JHEP **03**, 119 (2018), arXiv:1712.05150 [hep-th] (cit. on p. 3).

[23] I. Buchbinder, S. J. Gates, and K. Koutrolikos, "Integer superspin supercurrents of matter supermultiplets," JHEP **05**, 031 (2019), arXiv:1811.12858 [hep-th] (cit. on p. 3).

[24] E. A. Ivanov, "On the Geometric Meaning of the $N = 1$ Yang-Mills Prepotential," Phys. Lett. B **117**, edited by P. Petiau and M. Porneuf, 59 (1982) (cit. on p. 3).

[25] E. A. Ivanov, "Intrinsic Geometry of the $N = 1$ Supersymmetric Yang-Mills Theory," J. Phys. A **16**, 2571 (1983) (cit. on p. 3).

[26] I. L. Buchbinder, S. J. Gates, and K. Koutrolikos, "Superspace first order formalism, trivial symmetries and electromagnetic interactions of linearized supergravity," JHEP **09**, 077 (2021), arXiv:2107.06854 [hep-th] (cit. on pp. 3, 4, 16).

[27] K. Koutrolikos, "Superspace first-order formalism for massless arbitrary superspin supermultiplets," Phys. Rev. D **105**, 125008 (2022), arXiv:2204.04181 [hep-th] (cit. on pp. 3, 4, 16).

[28] I. Buchbinder and T. Snegirev, "Lagrangian formulation of free arbitrary N-extended massless higher spin supermultiplets in 4D, AdS space," (2020), arXiv:2009.00896 [hep-th] (cit. on p. 5).

[29] S. J. Gates Jr., M. T. Grisaru, and S. Penati, "Holomorphy, minimal homotopy and the 4-D, N=1 supersymmetric Bardeen-Gross-Jackiw anomaly," Phys. Lett. B **481**, 397–407 (2000), arXiv:hep-th/0002045 (cit. on p. 7).

[30] S. J. Gates Jr., M. T. Grisaru, M. E. Knutt, S. Penati, and H. Suzuki, "Supersymmetric gauge anomaly with general homotopic paths," Nucl. Phys. B **596**, 315–347 (2001), arXiv:hep-th/0009192 (cit. on p. 7).

[31] S. J. Gates Jr., M. T. Grisaru, M. E. Knutt, and S. Penati, "The Superspace WZNW action for 4-D, N=1 supersymmetric QCD," Phys. Lett. B **503**, 349–354 (2001), arXiv:hep-ph/0012301 (cit. on p. 7).

[32] D. Jain and W. Siegel, "N=2 Super-Yang-Mills Theory from a Chern-Simons Action," Phys. Rev. D **86**, 125017 (2012), arXiv:1203.2929 [hep-th] (cit. on p. 7).

[33] A. Davgadorj, U. Lindström, and R. von Unge, "New techniques for gauge theories in projective superspace," Phys. Rev. D

**108**, 066004 (2023), arXiv:2306.10399 [hep-th] (cit. on p. 7).

[34] S. Gates, M. T. Grisaru, M. Rocek, and W. Siegel, *Superspace Or One Thousand and One Lessons in Supersymmetry*, Vol. 58 (1983), arXiv:hep-th/0108200 (cit. on pp. 8, 9).

[35] I. L. Buchbinder and S. M. Kuzenko, *Ideas and methods of supersymmetry and supergravity: Or a walk through superspace* (1998) (cit. on p. 9).

[36] S. J. Gates Jr. and K. Koutrolikos, "On 4D, $\mathcal{N} = 1$ massless gauge superfields of arbitrary superhelicity," JHEP **06**, 098 (2014), arXiv:1310.7385 [hep-th] (cit. on p. 12).

[37] B. de Wit and D. Z. Freedman, "On Combined Supersymmetric and Gauge Invariant Field Theories," Phys. Rev. D **12**, 2286 (1975) (cit. on p. 14).